# EBCD: A Routing algorithm based on bee colony for energy consumption reduction in wireless relay networks

Arash Ghorbannia Delavar[1] and Elham Javiz[2]

[1]Department of Computer Engineering and Information Technology,
Payam Noor University, Tehran, Iran
a_Ghorbannia@pnu.ac.ir
[2] Department of Computer Engineering and Information Technology,
Payam Noor University, Tehran, Iran
javiz.ssg@gmail.com

## ABSTRACT

*One of the important issues in wireless networks is the Routing problem that is effective on system performance, in this article the attempt is made to propose a routing algorithm using the bee colony in order to reduce energy consumption in wireless relay networks. In EBCD algorithm, through combined of energy, distance and traffic parameters a routing algorithm for wireless networks is presented with more efficiency than its predecessor. Applying the bee colony method would allow the placement of the parameters under conventional conditions and to get closer to a mechanism with a better adaptability than that of the existing algorithm. According to the parameters considered, the proposed algorithm provides a fitness function that can be applied as a multi-hop. Unlike other algorithms of its kind this can increase service quality based on environmental conditions through its multiple services. This new method can store the energy accumulated in the nodes and reduce the hop restrictions.*

## 1. INTRODUCTION

Multi hop wireless systems have the potential to offer more coverage and capacity over single-hop radio access systems. There are a number of different types of multi hop wireless networks, notably the ad hoc networks, sensor networks, and wireless mesh networks. Each one of these network types has different characteristics that results having different systems in design and routing protocols. Another type of the multi hop wireless networks that is a subject of focus is based on relay architecture [2]. In the recent years, many studies have been conducted on the wireless communication via relays. The relays have their essential input in future communication networks. Relay-based systems are typically composed of small form factor low-cost relays, which are associated with specific base stations (BSs). In general , the relays could be used in the initial layers of the network to provide more coverage and cover vast areas at lower cost than that of the  BS; they can also be used in providing increased capacity in more developed networks and cover the holes such as the shaded areas of the buildings. The IEEE 802.16 standard is developed to provide broadband wireless access in 4G systems. Its physical layer adopts the OFDMA technique, where a base station (BS) can communicate with multiple mobile stations (MSs) simultaneously via orthogonal channels. Since the typical PMP (point-to-multi-point) operation could face several problems such as holes coverage and network congestion at the BS, 802.16j has developed a standard to support relay mode function and solve these problems in 802.16 systems [1]. IEEE 802.16j network includes one Base Station (BS), multiple Relay Stations (RSs) and Mobile Stations (MSs). All MSs are under the BS's signal coverage and RSs are located in the BS's coverage boundary in order to relay data between MSs and the BS. There are two types of relays: transparent and non-transparent. The transparent relays are used in increasing the capacity within the BS coverage area. These relays have low complexity and serve in topologies up to two hops. The non-transparent relays are used in increasing the coverage areas. These relays have different levels of complexity and are used in topologies with more than two hops. Since the 802.16j relay networks have multi hop paths between the BS and MS, naturally issues regarding routing and path management rise. Routing and path management functions manage the Multi-hop paths between the BS and MS. Routing is based on a tree structure where the BS is root and the MSs are the leaves. Although routing is tree-based the decisions could be made regarding which RS is associated with which MS [2].

Generally there exist three scenarios in IEEE 802.16j relay networks that include [13]:

1. The MS is connected directly to BS

2. The MS is connected to BS through a transparent relay

3. The MS is connected to BS through one or more non-transparent relay

As shown in Figure 1, this newly introduced routing algorithm works on all three scenarios. In practice, much research have worked in the two-hop relays but have not supported the ability to increase the capacity of relays, although the two-hop cannot full fill the objective of having mobile users. This point is considered in the proposed algorithm here.

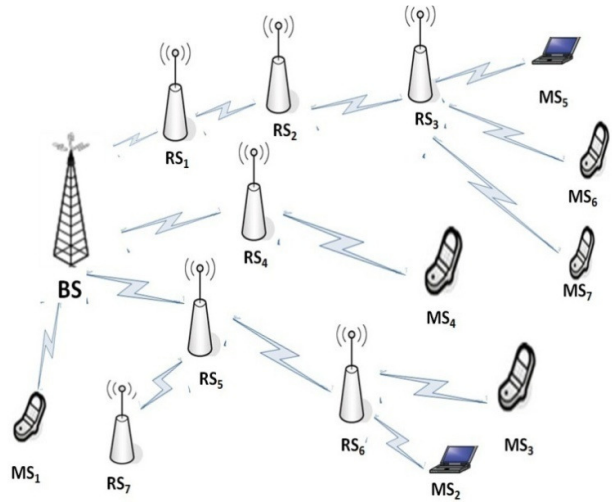

Figure 1: An example of usage scenario

## 2. RELATED WORKS

So far, several algorithms for routing in Wireless Multi-Hop Relay networks are presented. C. Hong et.al [6] has proposed a routing algorithm based on linear programming. In this algorithm, the routing is based on the total time to send and receive data from source to destination. K. Wang et.al [7] has presented a route selection based on the optimum throughput. This algorithm is focused on two-hop networks. D.J. Son et.al [8] has offered a route selection scheme based on Spectral Efficiency and load traffic for IEEE802.16j networks. K.P. Shih et.al [9] has introduced a new parameter called SEL for evaluating the optimal path. SEL considers the Spectral Efficiency parameters and link load. In their article the focus is only on the Two-Hop Relay Networks [4].

## 3. THE BEE COLONY ALGORITHM

Considering the use of biological ideas in routing algorithms is an almost new concept. The idea of these algorithms was initialled in wired networks and then extended to wireless networks. One of these ideas is inspired from bees search in routing that is copied in wireless networks [3], [14], [16], [17]. In a bee colony a strange phenomenon is observed which is characterized by the behavioural manner of every individual bee under different circumstances. Although every bee follow simple rules, their collective behaviour is very intelligent; therefore, this aspect of the bee colony is applied in wireless relay networks.

### 1.3. Characteristics of a bee colony

The study conducted upon bee colonies and their inter communication system by Karl Frisch has revealed many aspects of their communicative system. Every bee return to the hive conveys for information on distance, direction and quality of food to the other bees through specific dance [15]. By applying this mechanism the bees in the hive begin their journey to collect their food. Bee colonies have unique characteristics like:

- appropriate and efficient distribution of bees among different food sources

- the appropriate division of labor

- testing and estimating the quality of food by every bee and sending the appropriate number of bees to the food source based on its quality through the dance

- no central control system

- decision making about the suitability of a food source considering the energy required for food collection

In a bee colony, bees have the same behavioral structure, but every bee according to colony's needs is equipped with different skills over time. For example, a bee may be required to collect food and in other time phase it might be required for storing the food. This type of behavior causes the colony to be flexible against the environmental changes. A very intriguing and subtle point in the duty distribution system in a bee hive is that the bee in getting information from the other bees in a close circuit. This fact prevents that sudden swarming all the bees don't bring an influx to the best source, a common effort for the colony's welfare.

## 4. PROBLEM DEFINITION

The data transmission among stations in the Wireless relay networks is synchronized on frame basis. Every frame consists of several time slots defined as a basic time unit for transmission in the system [6]. To every MS a burst is assigned for data transmission to its uplinks that include many time slots and multiple slots associated with the MCS (Modulation and Coding Schema) level. Based on the MCS level the amount of transferable data (in bits) is determined, in each slot [1]. Energy consumption during each frame for data transmission between $MS_i$ and $RS_j$ is obtained through the following equation [1]:

$$(1)$$

$$E_{MR} = \left( \left\lceil \frac{d_i}{D(k)} \right\rceil \times \tau \right) \times \frac{10^{\frac{\delta(k)}{10}} (B \cdot N_0 + I(i,j)) \cdot L(i,j)}{G_i \cdot G_j}$$

Where, $d_i$ is the number of bits in a request at the current frame, $D(k)$ is the amount of bits transmitted by a slot, $\tau$ is the length of a slot (in seconds), $B$ is effective channel bandwidth, $G_i$ is antenna gains at $MS_i$, $G_j$ is antenna gains at $RS_j$, $No$ is the thermal noise level.

$I(i,j)$ is the interference of the simultaneously transmitted, calculated through the following equation [1]:

$$I(i,j) = \sum_{i \neq i'} P(i', j)$$

$$(2)$$

When $MS_i$ sends data to an $RS_j$ by using the transmission power $P_i$ and MCS level $M_i$, the received signal power $P(i,j)$ at $RS_j$ is presented as [1]:

$$P(i,j) = \frac{G_i \cdot G_j \cdot P_i(M_i)}{L(i,j)}$$

$$(3)$$

Where, $L(i,j)$ is the path loss from $MS_i$ to $RS_j$, which is calculated based on SUI (Stanford university interim) path loss model [11], [12]. There exists a direct relation between distance increase and the amount of BS path- loss.

## 5. PROPOSED ALGORITHM

The flexible nature and the ability to adopt with new environment and the bee colony's decision making features regarding a specific food source-with respect to the quantitative and qualitative aspects of the food source- and the energy needed to reach to the source are considered here in developing a routing algorithm with the less energy consumption objective.. In this article the up Link traffic is evaluated. The Down Link traffic is calculated in a similar approach where the details are ignored. The Non-transparent relay for communication support over a two-hop and transparent relay for increased capacity within the two-hop are considered here and the BS is evaluated as a specific RS.

In this algorithm, first an initial population of random solutions is considered; then for each solution all useful paths are evaluated by F fitness function and through Elitism Selection method a route is chosen, this process is repeated until the maximum number of allowable steps are obtained; then the objective function value of each of the solutions is evaluated through the following equation:

$$p_i = \frac{f_i}{\sum_{k=1}^{n} f_k}$$

$$(4)$$

Here, the algorithm steps are repeated until the stopping condition is fulfilled .As mentioned before, in order to assess the routes of every solution the fitness function F is applied which is obtained by the following equation:

$$(5)$$

$$F = E + T + \frac{1}{Dist}$$

The parameters used in the fitness function consist of:

A- $E$ is the total amount of energy consumed for data transmission from MS to the BS

$$E = E_{MR} + E_{RR} + E_{RB} \qquad (6)$$

Where, $E_{MR}$ is the amount of energy consumed for data transmission from MS to the RS, $E_{RR}$ is the amount of energy consumed for data transmission from RS to the RS, $E_{RB}$ is the amount of energy consumed for data transmission from RS to the BS, $E_{MR}$, $E_{RR}$ and $E_{RB}$ are calculated through equation (1)

B- $T$ is the data transmission cost or traffic that is obtained through the following equation:

$$T = \frac{d_i}{BW} \qquad (7)$$

C-Dist is the distance parameter considered as equal to the received signal power in this realm. The less the distance the more signal receiving power. This parameter is calculated through equation (3)

Chart 1shows the flowchart of this proposed routing algorithm which is based on the bee colony optimization algorithm.

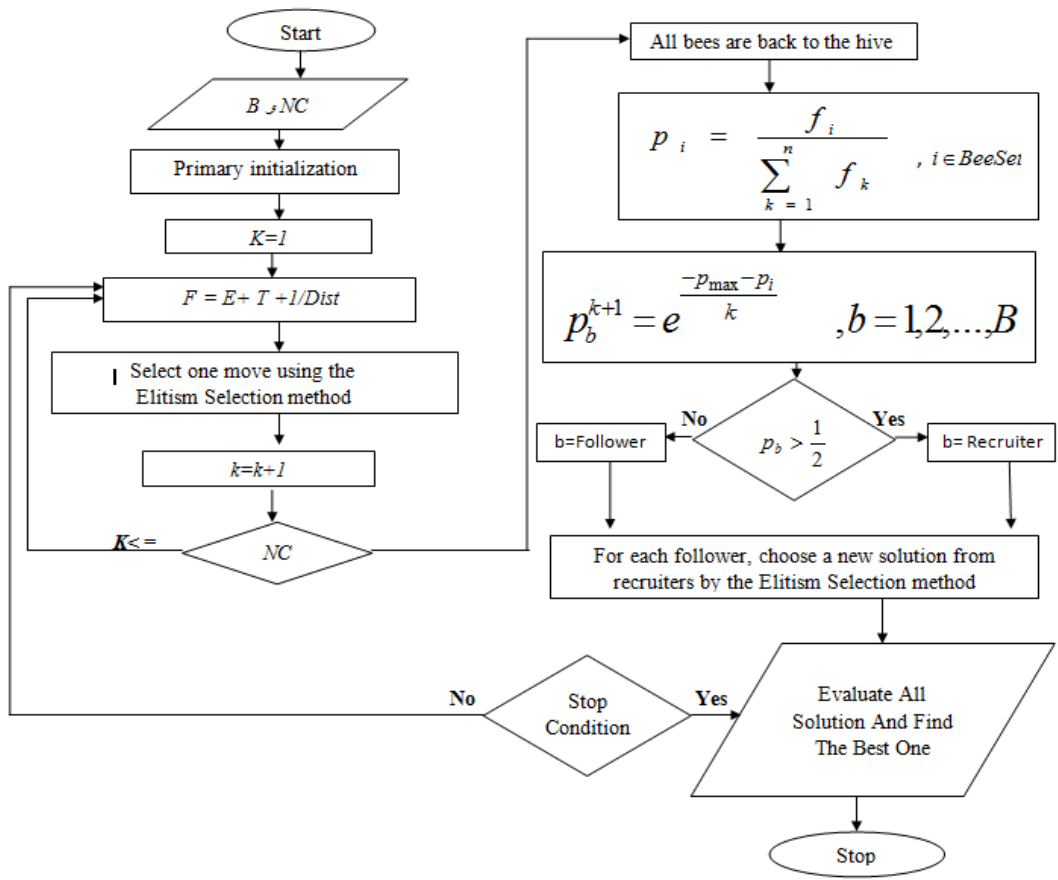

Chart1. Proposed Algorithm Flowchart

## 6. SIMULATION

The MATLAB software is used here in order to simulate the proposed algorithm. In this conducted simulation nodes are deployed in the environment randomly and the run time is 10 seconds (2000 frames).The simulation parameters are presented in Table 1.

Table1: Simulation Parameters

| Parameter | Value |
|---|---|
| Channel Bandwidth(BW) | Rand[3.5,10]MHZ |
| $d_i$ | Rand[900,2000] bits/frame |
| Antenna height | BS:30m,RS:10m,MS:2m |
| ($NO$)Thermal Noise | -100dBm |
| $p_i^{max}$ | 1000mw |
| BS,RS antenna gain | Rand[5,20]dB |
| MS antenna gain | Rand[1,10]dB |
| d | Rand[200,2000]m |
| Frame Duration | 5ms |
| Slots Per Frame | 48 |

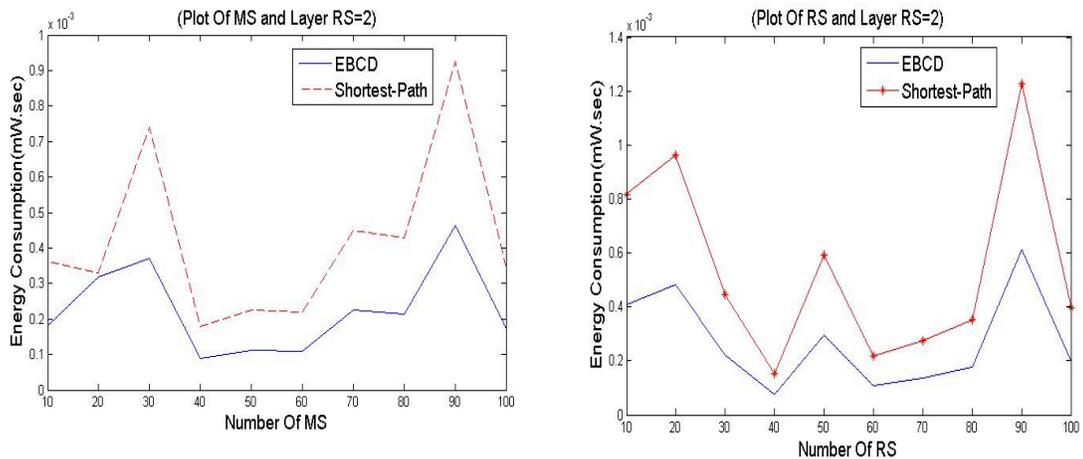

a)  In the different number of MSs and 10 RSs    b) In the different number of RSs and 100 MSs

Figure2: The energy consumption in the Network with 3 -hops

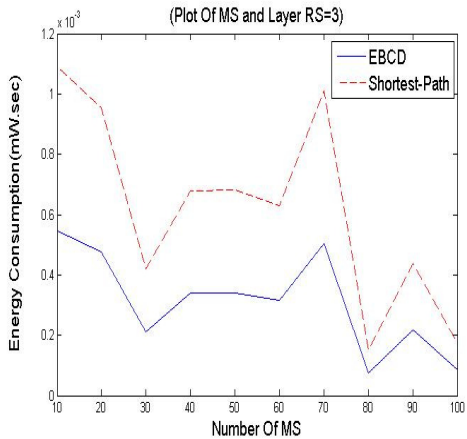 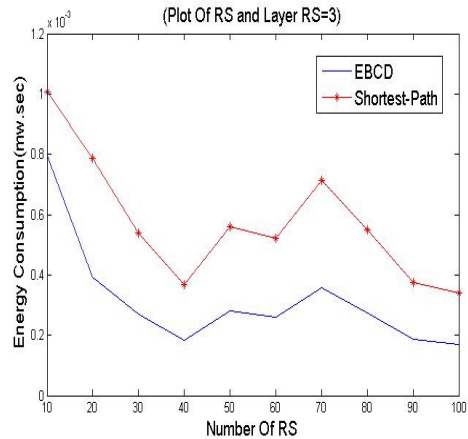

a) In the different number of MSs and 20 RSs

b) In the different number of RSs and 100 MSs

Figure3: The energy consumption in the Network with 4 -hops

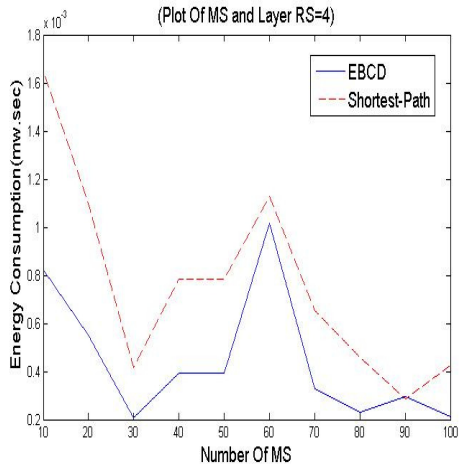 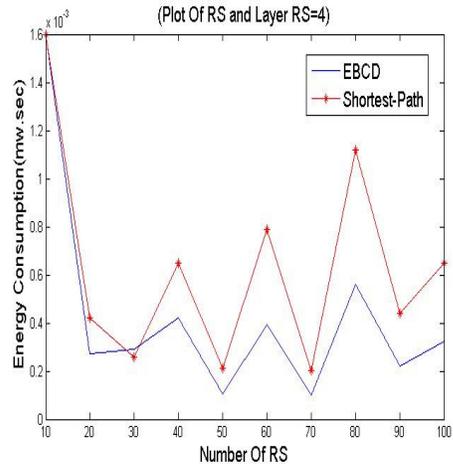

a) In the different number of MSs and 30 RSs

b) In the different number of RSs and 100 MSs

Figure4: The energy consumption in the Network with 5 -hops

This proposed algorithm is compared with the Dijkstra's shortest-path algorithm. Here three scenarios are considered for Comparisons. In the first, the network topology is of 3 hop maximum i.e.2 relay layers between MS and BS. The total energy in each frame is calculated by having different number of MSs, where there are 10 RSs. As shown in Figure 2(a). The y-axis is presented in an exponential scale. Indicating that here, the energy saving is approximately optimized by 4.8% in comparison with that of the Dijkstra's shortest-path algorithm. Then the total energy with respect to different numbers of RSs and 100 MSs in Figure 2(b) are calculated. Here a 5% optimization in energy consumption is. The topology used in the second scenario is evident up to 4 steps. According to Figure 3(a) the total energy consumed with a number of different MSs and 20 RSs in this proposed algorithm is reduced up to 5.1%. Figure 3(b) indicates that with a variable number of RSs and 100 MSs the total energy consumed in EBCD is approximately improved by 3.9%. In the third scenario, the steps are of 5 hop maximum. The total energy consumed in this algorithm with variable number of MSs and 30 RSs in figure 4(a) is about 4% less than that of the Dijkstra's shortest-path. In figure 4(b) with 100 MSs and a variable number of RSs the total energy consumed in EBCD is approximately improved by 3.4%. In general the Numerical results of the simulations indicate that this proposed routing algorithm is efficient in terms of reducing energy consumption.

# 7. CONCLUSION

In this article a new algorithm for routing in the relay networks using the properties of the bee colony with an emphasis on reducing energy consumption is presented. Unlike most works in this field that have been working with the two-step, here the number of hops is extended to 5. For the measuring function and evaluating the effective environmental parameters of energy, the distance and traffic are considered. This Proposed algorithm compared to other known algorithms called Dijkstra's Shortest path algorithm indicate that EBCD is about 4.3% more efficient than the Dijkstra's shortest-path in energy consumption.

**Authors**

**Arash Ghorbannia Delavar** received the MSc and Ph.D. degrees in computer engineering from Sciences and Research University, Tehran, IRAN, in 2002 and 2007. He obtained the top student award in Ph.D. course. He is currently an assistant professor in the Department of Computer Science, Payam Noor University, Tehran,  IRAN. He is also the Director of Virtual University and Multimedia Training Department of Payam Noor University in IRAN. Dr. Arash Ghorbannia Delavar is currently editor of many computer science journals in IRAN. His research interests are in the areas of computer networks, microprocessors, data mining, Information Technology, and E-Learning.

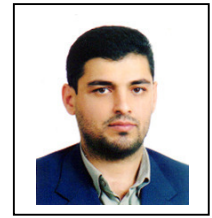

**Elham Javiz** received the B.Sc. in computer engineering from Azad University, najafabad, IRAN, in 2004. She is M.Sc. student in computer engineering in Payam Noor University. Her research interests include computer networks, wireless communication, wireless multi hop networks and Bee colony algorithm.

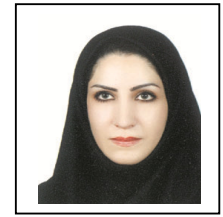